\documentclass[12pt,a4paper,oneside,noupper]{article}
 \usepackage{amsmath}
  \usepackage{amsfonts}
\usepackage{amssymb}
 \usepackage{graphicx}
  
 \newcommand{\singlefig}{.75\textwidth}
 
  \newcommand{\ii}{\mathrm{i}}
\begin{document}

\date{15 Jul 2003}
\title{Effect of base-pair inhomogeneities on charge
transport along DNA mediated by twist and radial polarons}

\author{
{\bf F Palmero, JFR Archilla} \\ ETS Ingenier\'{\i}a
Inform\'atica. Universidad de Sevilla.  \\ Avda Reina Mercedes
s/n, 41012-Sevilla, Spain \\ {\bf D Hennig} \\ Freie Universit\"at Berlin,
\\ Fachbereich Physik, Arnimallee 14,
14195-Berlin (Germany)\\{\bf FR Romero} \\ Facultad de
F\'{\i}sica. Universidad de Sevilla.  \\ Avda Reina Mercedes s/n,
41012-Sevilla, Spain}
\maketitle

\begin{abstract}

Some recent results for a three--dimensional, semi--classical,
tight--binding model for DNA show that there are two types of
polarons, named radial and twist polarons, that can transport
charge along the DNA molecule. However, the existence of two
types of base pairs in real DNA, makes it crucial to find out if
charge transport also exist in DNA chains with different base
pairs. In this paper we address this problem in its simple case,
an homogeneous chain except for a single different base pair,
what we call a base-pair inhomogeneity, and its effect on charge
transport. Radial polarons experience either reflection or trapping.
However, twist polarons
are good candidates for charge transport along real DNA. This
transport is  also very robust with respect to weak parametric
and diagonal disorder.

\end{abstract}

\section{Introduction}
Charge transport in biomolecules is of 
  relevant
in many biological
functions. 
 Moveover,
 electronic transport plays a well known and
fundamental role in 
 many
 of biological processes, such as
nervous conduction, photosynthesis, cellular respiration and redox
reactions. Recently, DNA--mediated charge transport has focused
much attention 
  with
   experiments that have been conducted to
determine its conductive properties, and theoretical studies have
addressed 
  the possibility of charge transport and its efficiency
 \cite{Gas97}. The importance of charge transport through DNA is
significant
   because some mutations in living
systems and radical migrations are critical issues in
carcinogenesis studies and may yield insight into damage prevention
or repair 
 processes
\cite{Boi02, Bur98, Fri01}. Moreover, charge transport has
been demonstrated to proceed within HeLa cell nuclei \cite{Meg01}
as well as in the nucleosome core particles \cite{Meg00}, and can
provide a 
  practical method of genetic screening for
known gene sequences and 
  an alternative method to
hybridization based arrays \cite{Mar02}. Material scientists have
thought that DNA has a fundamental physical interest for the
development of DNA--based molecular technologies, as it possesses
ideal structural and molecular recognition properties for use in
self assembling nanodevices with a definite molecular
architecture \cite{Mao00}.

Photochemical electron transfer experiments have been
carried out with  femtosecond
 resolution, and have shown, with the
precise control of the DNA sequences, a fast time scale ($\leq 5 $
ps) process, attributed to charge mobility along an unperturbed
double strand, and a slower process ($\leq 75 $ ps), that they
believe reflects charge hopping between perturbed domains along
the DNA strand \cite{Wan99}.

Experimental results have explained that DNA acts as a linear chain
with overlapping $\pi$ orbitals located at the stacked base pairs and
its conductivity is very sensitive to disruption, caused by
base--pair mistakes or interactions with proteins, in the base--pair
stack. Depending on the base pair sequences the transfer can be slowed
or inhibited \cite{Wan00}.

Moreover, the robust, malleable one--dimensional structure of DNA can
be used to design electronic devices based on biomaterials
~\cite{Ratner99,Fink99,Tran00,Braun98,Porath00}.


The structure of the bent double helix $\lambda$-DNA can be
modelled by a network of oscillators taking into account
deformations of the hydrogen bonds within a base pair and twist
motions between consecutive base pairs. The three--dimensional
semi--classical tight--binding model for DNA that was first
proposed in Ref.~\cite{HAA03,AHA03} makes also the assumption that
the electron motion is predominantly influenced by vibrational
modes of the double helix. The nonlinear interaction between the
electron and the vibrational modes is responsible of the formation
of polarons or electron-vibron breathers. These are localized
excitation patterns that can be static, producing charge
localization, or mobile, bringing about charge transport. It
has been considered some other models with inhomogeneities
\cite{Mut92}.

A variant of this model has been proposed in Ref.~\cite{PAHR03},
which lifts the restriction of the previous one, that the
perturbation of the spatial variables were small enough to perform
a linear approximation in the dynamical equations. Two types of
polarons appear depending on the parameter  $\alpha$, that
describes the coupling between the transfer integral and the
distance between nucleotides, and the form of activation: radial
polarons and twist polarons. For radial polarons, the deformations
associated by the electronic charge affect mainly to the radial
variables, and for twist polarons, to the angular ones. In the
last case, the moving polaron is slower than in the previous one,
but it is more robust with respect to the introduction of parametric
disorder. See reference above for details and Ref.~\cite{zhang02}
for twist polarons in other model.

In this paper we consider  the movement of
polarons along DNA chains made out of different base pairs.
The most simple one consists of homogeneous DNA chains except for
a single different base pair, which we call base-pair inhomogeneity.

We have found that when charge transport is mediated by means of a
twist polaron, this moving carrier can transport charge
along a base-pair inhomogeneity in a very efficient way. Moreover, charge
transport persists under the introduction of a high degree of parametric
disorder, that represents the inherent disorder in the surrounding
medium and the inhomogeneous distribution of counterions along the
DNA duplex~\cite{BC01}. On the other hand, radial polarons are either
reflected or trapped by the base-pair inhomogeneity, making them poor candidates
for charge transport along heterogeneous DNA at least in the framework of the
tight-binding models.

\section{DNA model}
\subsection{Description}
We consider a variant~\cite{PAHR03} of the model Hamiltonian
for charge transport along DNA
that was first proposed in ~\cite{HAA03}. This variant consists of
considering the full dynamical equations of the system instead
of performing linear approximations based on the assumption that
the deformations of the molecule were small.

These models were designed to implement the basic characteristics
of the DNA double helix structure needed for a proper description
of the charge transport dynamics. The starting point is the {\em
twist-opening} model ~\cite{Barbi98,Barbi99a,Barbi99b} that has
taken into account the helicoidal structure of DNA and the
torsional deformations induced by the opening of the base pairs.
The bases are considered as single nondeformable objects. The
helicoidal structure of DNA is described in a cylindrical
reference system and the n--th base pair has two degrees of
freedom, namely $(r_n,\phi_n)$, where $r_n$ represents the radial
displacement of the base pair from the equilibrium value $R_0$,
and $\phi_n$ represents the angular deviation from equilibrium
angles with respect to a fixed external reference frame.  As we
are interested in base pair vibrations and not in acoustic
motions, the center of mass of each base pair is fixed, $i.e.$,
the two bases in a base pair are constrained to move symmetrically
with respect to the molecule axis. Moreover, the distances between
two neighboring base pair planes will be treated as fixed, because
in the axial direction DNA is less deformable than within the base
pair planes~\cite{Barbi98}.

\begin{figure}
  \begin{center}
 \includegraphics[width=\singlefig]{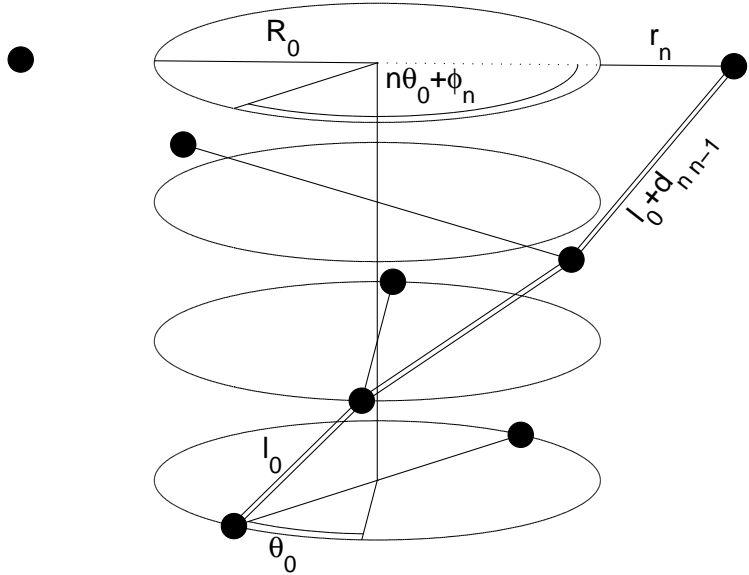}
  \end{center}
  \caption{Sketch of the model. Filled circles  represent  bases. The variables
  used in the text are displayed.}
  \label{model}
\end{figure}

A sketch of the helicoidal structure
of the DNA model is shown in Fig.~\ref{model}.

The transversal displacements of the base pairs are deformations
of the H-bonds and the angular twist and the radial vibrational
motion evolve independently on two different time scales, then
they can be considered as decoupled degrees of freedom in the
harmonic approximation of the normal mode
vibrations~\cite{Cocco00}.

\subsection{Hamiltonian}
The Hamiltonian for the electron transport along a strand of DNA
is given by $
\widehat{H}=\widehat{H}_{el}+\widehat{H}_{rad}+\widehat{H}_{twist}$,
where $\widehat{H}_{el}$ corresponds to the part related with the
particle charge transport over the base pairs,$\widehat{H}_{rad}$
describes the dynamics of the H-bond vibrations and
$\widehat{H}_{twist}$ is the part corresponding to the dynamics of
the relative twist angle between two consecutive base pairs.
 This electronic part is described by a tight--binding system of the form:
\begin{equation}
\widehat{H}_{el}=\sum_nE_n|n \rangle \langle n|-V_{n-1,n}|n-1
\rangle \langle n|-V_{n+1,n}|n+1 \rangle \langle n|\,,
\end{equation}
where $|n \rangle $ represents a localized state of  the charge
carrier at the $n^{th}$ base pair. The quantities $\{V_{n,n-1}\}$
are the nearest--neighbor transfer integrals along  base pairs,
and $\{E_n\}$ are the energy on--site matrix elements. A general
electronic state is given by $|\Psi\rangle=\sum_n
c_n(t)|n\rangle$, where $c_n(t)$ is the probability amplitude of
finding the charged particle in the state $|n\rangle$. The time
evolution of the $\{c_n(t)\}$ is obtained from the
Schr$\ddot{o}$dinger equation $\ii \hbar(\partial \Psi/ \partial
t)=\widehat{H}_{el}|\Psi\rangle$

The nucleotides are large molecules and they move much more slowly
than a charged particle, then the lattice oscillators can be
described classically and $\widehat{H}_{rad}$ and
$\widehat{H}_{twist}$ are, {\em de facto}, classical Hamiltonians.
For homogeneous chains they are  given by (omitting the hat over them):
\begin{eqnarray}
H_{rad} &=& \sum_n \left[ \frac{1}{2M} (p^r_n)^2+\frac{M
\Omega^2_{r,n}}{2}r_n^2 \right],\\  H_{twist} &=& \sum_n  \left[
\frac{1}{2J} (p^{\phi}_n)^2+\frac{J \Omega_\phi^2}{2}\
(\phi_n-\phi_{n-1})^2 \right],
\end{eqnarray}
where $p^r_n$ and $p^{\phi}_n$ are the conjugate momenta of the
radial and angular coordinates,respectively. In these expressions
$M$ is the mass of each base pair ($M=2m$,being $m$ an average
estimation of the nucleotide mass), $J=M\,R_0^2$ is the moment of
inertia of each base pair and $\Omega_{r,n}$ is the linear radial
frequency which is proportional to the strength of the hydrogen
bonds. Indeed, for an homogeneous chain all these frequencies are
equals, that is, $\Omega^2_{r,n}=b_n\Omega^2_{r}$ with
$b_n=1\,\forall n$.

 In this paper we are interested in the study of charge transport
along DNA strands where all the base pairs are of the same
type, say A-T (or C-G), except only one of them which is of a
different type, say C-G (or A-T). We can suppose that the ratio between the
elastic constants of bonds in a C-G base pair and an A-T base
pair is $3/2$ because the first involves three hydrogen bonds and the second two
of them. Then, as in  Ref.~\cite{Sa91} we take  $b_n=0.8$ for an
A-T base pair, and $b_n=1.2$ for a C-G base pair.
 $\Omega_\phi$ is the linear twist frequency, and we represent by
$\theta_{n,n-1}=(\phi_n-\phi_{n-1})$, the deviation of the
relative angle between two adjacent base pairs from its
equilibrium value $\theta_0$.

In general, the ionization potential of different nucleotides
differs an amount of 0.2-1.0 eV \cite{Bru02}. In our model, this
implies different values of the on site energies $E_n^0$ for each
base pairs. In this work, we have focused in geometrical effects
due to the stretching of the chain over the charge, and we have
considered $E_n^0$ independent on the type of base.

 The electronic part of the Hamiltonian, $\widehat{H}_{el}$, has a
dependence on the structure variables $r_n$ and $\phi_n$
through the dependence of the matrix elements $E_n$ and
$V_{n,n-1}$ on them. The energy on--site matrix elements are given
by~\cite{KAT98} $E_n = E_n^0+k r_n$, expressing the modulation of
the on--site electronic energies $\{E_n^0\}$ by the radial
deformations of the base pairs. The transfer matrix elements
$V_{n,n-1}$, which are responsible for the transport of the
electron along the stacked base pairs, are assumed to depend on
the distances between two consecutive bases along a strand
$d_{n,n-1}$ as
 $V_{n,n-1} = V_0(1-\alpha d_{n,n-1})$,
where $\alpha$ is a parameter that describes the influence of the
distance between nucleotides, and the later is determined by
\begin{eqnarray}
\label{distance}
\nonumber d_{n,n-1} & = & [a^2+(R_0+r_n)^2+ (R_0+r_{n-1})^2-
\\
& & 2(R_0+r_n)(R_0+r_{n-1})
\cos(\theta_0+\theta_{n,n-1})]^{1/2}-l_0,
\end{eqnarray}
with
 $
l_0=(a^2+4R_0^2 \sin^2(\theta_0/2))^{1/2},
 $
where $a$ is the vertical distance between two consecutive base
pairs.

 In this paper, as in Ref.~\cite{PAHR03},
 we have not used the expansion of this expression up
to first order around the equilibrium positions, as was performed
in Refs.~\cite{HAA03,AHA03}. This allow us to consider parameters that
allow larger deformations in the angular variables.

Realistic parameters for the DNA are given in
Refs.~\cite{Barbi99a,Stryer95}. We have considered: $a= 3.4 \AA$,
$m=300$ amu, $R_0= 10 \AA$,$\Omega_r=8 \times 10^{12}$ s$^{-1}$,
$\Omega_{\phi}=9 \times 10^{11}$ s$^{-1}$. Ab initio calculations
find that between adjacent nucleotides, the transfer integral is
in order of 0.1-0.4 eV \cite{Sug96} . Although this transfer
integral is different for each pair of different nucleotides, we
will consider the same value for all neighboring cases $V_0=0.1$
eV, a supposition widely used and can be valid in order to
reproduce ab initio results and experiments \cite{Cun02}.

 We scale the time according to $t \rightarrow \Omega_r t$, and we
introduce the dimensionless quantities:
 $\tilde{r}_{n}=r_n\,(M \Omega_{r}^{2}/V_0)^{1/2}$,
 $\tilde{k}_{n}=k_n/(M\Omega_{r}^{2}V_0)^{1/2}$,
 $ \tilde{E_n}=E_n/V_0$,
 $\tilde{\Omega}=\Omega_{\phi}/\Omega_r$,
 $\tilde{V}=V_0/(J\,\Omega_{r}^2)$,
 $\tilde{\alpha}=\alpha\, (V_0/M\,\Omega_r^2)^{1/2}$,
 $\tilde{R}_{0}=R_0\,(M\,\Omega_r^2/V_0)^{1/2}$.

\subsection{Dynamical equations}
 Using the expectation value for the electronic contribution to the
Hamiltonian, the new classical Hamiltonian $\overline{H}=\langle \phi
|\widehat{H}|\phi\rangle /V_0$ is given in the scaled variables
 (omitting the tildes) by
\begin{eqnarray}
\nonumber \overline{H}&=&\sum_{n}\Big\{
\frac{1}{2}(\dot{r}_n^2+b_n r_n^2)+\frac{R_0^2}{2}[\dot{\phi_n^2}+\Omega^2(\phi_n-\phi_{n-1})^2]+
\\
 & &  (E_n^0+k r_n)|c_n|^2-(1-\alpha d_{n,n-1})(c_{n}^*c_{n-1}+c_{n}c_{n-1}^*) \Big \}.
\end{eqnarray}
The Schr\"odinger equation and the Hamiltonian equations lead to the
scaled dynamical equations of the system:
\begin{eqnarray}
\ii\,\tau\dot{c}_{n}&=&(E_n^0+k\,r_n)\,c_n\nonumber\\
\label{electron} &-&(1-\alpha\,d_{n+1,n})\,c_{n+1}
-(1-\alpha\,d_{n\,n-1})\,c_{n-1},
\\  \nonumber \ddot{r}_{n}&=&-b_n r_n-k\,|c_n|^2\ \\
& & -\, \alpha\, \left[\frac{\partial d_{n,n-1}}{\partial
r_n}(c_{n}^*c_{n-1}+c_{n}c_{n-1}^*)+\frac{\partial
d_{n+1,n}}{\partial r_n}(c_{n+1}^*c_{n}+c_{n+1}c_{n}^*) \right],
\label{radial} \\ \nonumber \ddot{\phi}_n
&=&-\Omega^2\,(2\phi_n-\phi_{n-1}-\phi_{n+1})\,
\\  & & -\,
  \alpha\,V\,\left[\frac{\partial d_{n,n-1}}{\partial \phi_n}
(c_{n}^*c_{n-1}+c_{n}c_{n-1}^*)+ \frac{\partial
d_{n+1,n}}{\partial \phi_n}(c_{n+1}^*c_{n}+c_{n+1}c_{n}^*)\right]
\label{angular}
\end{eqnarray}
where the quantity $\tau=\hbar\,\Omega_{r}/V_0$ measures the time
scale separation between the fast electron motion and the slow
bond vibrations. In the ordered case, with $E_n=E_0$, $\forall
{n}$, with the limit of $\alpha=0$  the set of coupled equations
represents the Holstein system, widely used in studies of polaron
dynamics in one-dimensional lattices. Also, for $\alpha=k=0$, and
random $E_n^0$, the Anderson model is obtained.

 The scaled parameters take the values  $\tau=0.053$,
$\Omega^2=0.013$, $V=2.5 \times 10^{-4}$, $R_0=63.1$ and
$l_0=44.5$. We fix the value  $k=1$ and consider the parameter
$\alpha$ as adjustable.

\section{Charge transport mediated by mobile polarons}

\subsection{Mobile polarons}
 The principal interest of this paper consists in the  study of
charge transport along the double strand by moving polarons in presence
of a base-pair inhomogeneity. We need first
to obtain localized stationary solutions of
Eqs.~(\ref{electron}-\ref{angular}). In Appendix A, we present the
procedure that we have followed for obtaining these stationary
solutions applied to different cases:  homogeneous chains,and
inhomogeneous chains.
 The polaron motion can be activated under certain conditions.
A systematic method to do this, known as the {\em pinning
mode}-method~\cite{CAT96}, consists of perturbing the (zero)
velocities of the ground state with localized, spatially
antisymmetric modes obtained in the vicinity of a bifurcation.
This method leads to moving entities with very low radiation, but
has the inconvenience of being applicable only in the neighborhood
of certain values of the parameters. An alternative is the
discrete gradient method~\cite{IST02}, perturbing  the (zero)
velocities of the stationary state $\{\dot{r_n}(0)\}$,
$\{\dot{\phi}_n(0)\}$ in a direction parallel to the vectors
$(\nabla r)_n=(r_{n+1}-r_{n-1})$ and/or $(\nabla
\phi)_n=(\phi_{n+1}-\phi_{n-1})$. Although this method does not
guarantee  mobility, it nevertheless proves to be successful in a
wide parameter range. We will denote the energy of the
perturbation as $\Delta K=\lambda^2/2$, where $\lambda$ is the
modulus of the vector used to perturb the system (parallel to
$\nabla r_n$ and/or $\nabla \phi_n$). This energy gives us the
difference of the energy between the moving polaron and the static
one, and it is usually called the activation energy.

\subsection{Homogeneous chains}

In this section we recall the basics results obtained in
Ref.~\cite{PAHR03}, see this reference for details. For
homogeneous chains in absence of diagonal disorder, $E_n^0=E^0,
\forall n$, and if the parameter $\alpha$ is small enough, it is
not possible to move the polarons by perturbing the angular
variables. Mobility can be accomplished only by perturbing through
the radial variables. However, for larger values of the parameter
$\alpha$ ($\alpha\gtrsim 0.01$), the polarons can only be moved by
perturbing the angular variables, i.e., perturbations of the
radial variables cannot activate mobility. Nevertheless, for
intermediate values of the parameter $\alpha$ ($0.005\lesssim
\alpha \lesssim 0.01$), the polarons can become mobile by
perturbing any set of variables, the radial or the angular ones.
The movement is rather different, when the polaron propagation is
activated by means of only radial perturbations, its
characteristics are similar to the radial movability regimen, and
likewise when it is activated by angular perturbations. A detailed
analysis of these different regimes can be found in the reference
above. In general, radial movability requires less energy and has
higher velocity than the angular one. Also, the limits of these
regimes are not exact, depending on the kinetic energy of the
perturbation.

\begin{figure}
\begin{center}
\includegraphics[width=\singlefig]{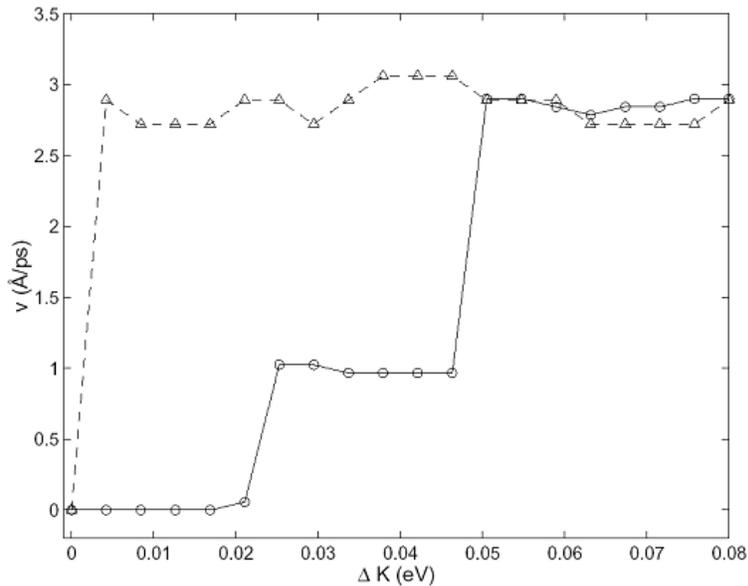}
\end{center}
\caption{Velocity of the polaron as a function of the energy
amount $\Delta K$ corresponding to the mixed regime  when the
polaron is activated by means of angular perturbations
($\alpha=0.01$). Circles and solid lines represent the disordered
case with $\Delta E=0.05$ and triangles and dashed line the
ordered case.}
 \label{meseta}
\end{figure}

 If an amount of  disorder in the on--site energies $E_n^0$ is
introduced, with random values $|E_n^0|<\Delta E$, we find that
moving polarons exist below a critical value  $\Delta E_{crit}$.
Beyond this value, polarons cannot be moved. In general, as shown
in Fig. \ref{meseta}, corresponding to the mixed regime where
polaron can be activated (in absence of disorder), by means of
both angular and radial perturbations, the mobility induced by
angular activation is more robust with respect to parametric
disorder. If disorder is high enough, radial perturbations that
could move polaron destroys it. Thus, it only can activated by
means of angular perturbations. In this case, the movement is
similar to the ordered case, the polaron has lower velocity, and
the activation energy is higher than in the radial movability
regime.

As shown in Fig.~\ref{meseta}, it can be appreciated that for a
polaron in the mixed regime, and if $\Delta E$ is high enough, it is
impossible to move it with radial perturbations, but only with
angular perturbations. In this situation, the movement is very
similar to the ordered case.

\subsection{DNA chains with a base-pair inhomogeneity}

 The studies of polarons in  homogeneous chains are only
applicable to synthetic DNA made out of a single type of base
pair. In real DNA, the two different base pairs, A-T and G-C,
combine in different ways constituting the genetic code. Thus, as
a first step in this direction, we consider a homogeneous
DNA-chain except for a single base pair of a different type.
Systems of this type are  easily synthesized and the numerical and
theoretical results could be compared with the experimental ones.
The outcome would help to determine the actual DNA parameters and,
therefore, the type of polarons that can be expected. There are
two different systems: a) a {\em hard--inhomogeneity}, that is, an
A-T DNA chain with a C-G inhomogeneity; b) a {\em soft
inhomogeneity}, i.e., a C-G DNA chain, with an A-T inhomogeneity.

 We activate the motion of a stationary polaron, using the discrete
gradient method, at a site about a hundred sites far away form the
location of the inhomogeneity.  Different types of moving polarons
can be obtained : radial, twist and mixed ones with different
energies, and we observe whether the polarons are reflected,
refracted or trapped.

\begin{itemize}

\item {\em Soft--inhomogeneity}

 In this case, radial polarons moving along a chain of C-G base
pairs are trapped by the soft A-T base pair as shown in
Fig.~\ref{trapping}. Trapping is caused by resonances between the
polaron and the stationary state centered at the inhomogeneity.

\begin{figure}
\begin{center}
\includegraphics[width=\singlefig]{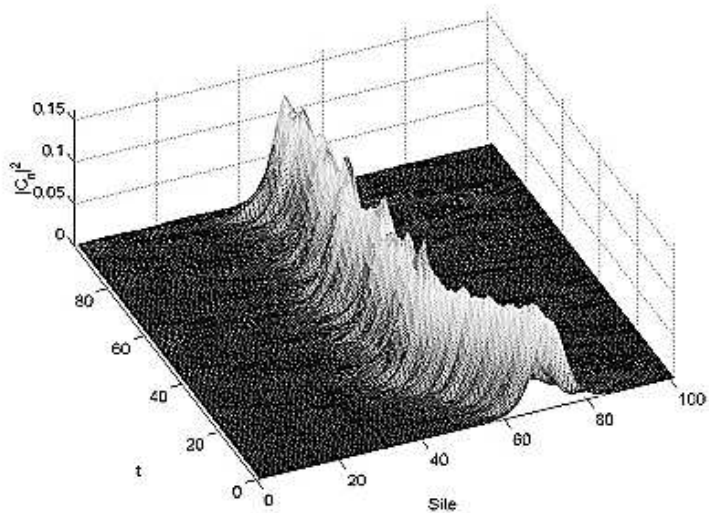} 
\end{center}
\caption{Soft inhomogeneity and radial polarons. Trapping
phenomenon due to the interaction between a moving radial polaron
in a C-G chain with an A-T base pair}
 \label{trapping}
\end{figure}

 However, when the polarons are activated by twist modes, the
moving polarons are always transmitted. The inhomogeneity acts as
a potential well, as shown in Fig.~\ref{tr1}. Note that the
polaron adopts briefly the shape of the ground state while passing
through the inhomogeneity.
\begin{figure}
\begin{center}
\includegraphics[width=\singlefig]{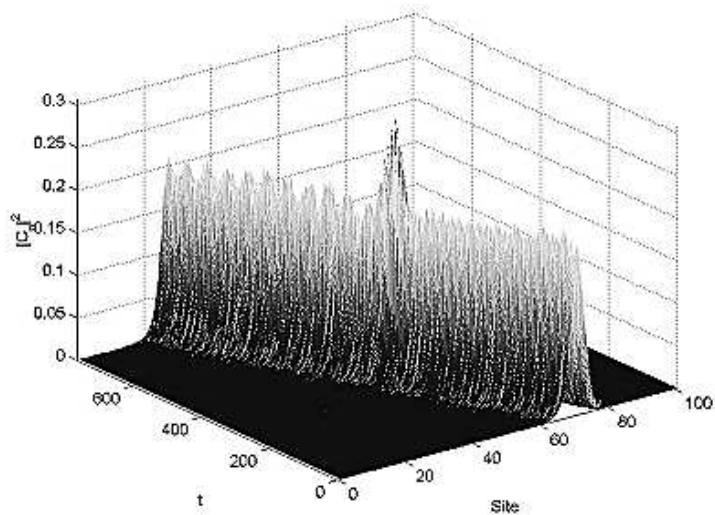} 
\end{center}
\caption{Soft inhomogeneity and twist polarons. Transmission
phenomenon due to the interaction between a moving polaron in a
C-G chain with an A-T base pair}
 \label{tr1}
\end{figure}

\item {\em Hard--inhomogeneity}

 In this case a radial polaron moving along a chain of A-T base
pairs interacts with the C-G inhomogeneity and  is always
reflected, as is shown in Fig.~\ref{reflection}. This phenomenon
is due to the impossibility of resonances, because the profiles of
the radial components of the polaron and the stationary one
located at the inhomogeneity are different.
\begin{figure}
\begin{center}
\includegraphics[width=\singlefig]{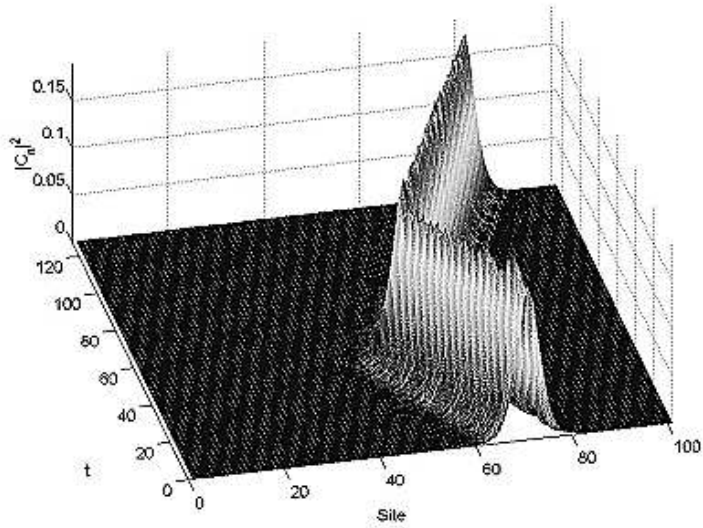} 
\end{center}
\caption{Hard inhomogeneity and radial polarons. Reflection
phenomenon due to the interaction between a moving polaron in an
A-T chain with a C-G base pair}
 \label{reflection}
\end{figure}

If the moving polaron is of the twist type, it is always
transmitted as shown in Fig.~\ref{tr2}. In this case the behavior
is similar to the movement of a particle through a potential
barrier, i.e, its velocity first decreases and then increases
until its entering value. Note that again the polaron adopts
briefly the shape of the ground state while passing through the
inhomogeneity.
\begin{figure}
\begin{center}
\includegraphics[width=\singlefig]{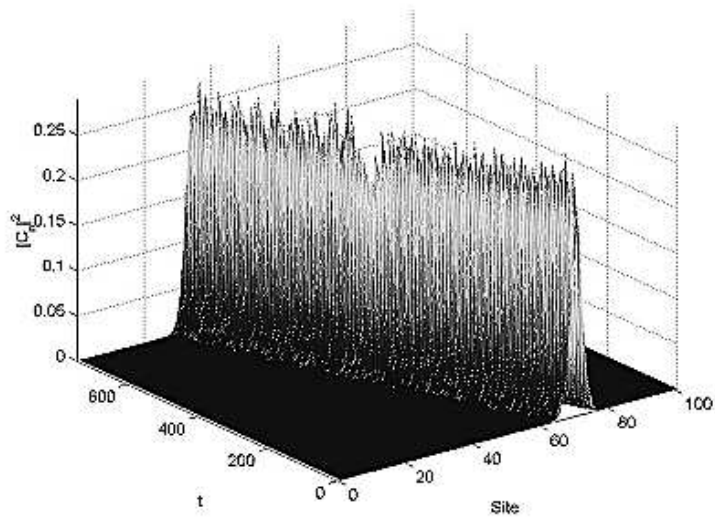} 
\end{center}
\caption{Hard inhomogeneity and twist polarons. Transmission
phenomenon due to the interaction between a moving polaron in an
A-T chain with a C-G base pair}
\label{tr2}
 \end{figure}

\end{itemize}

 In the mixed regime, the phenomena correspond exactly to the type
of polaron activated.

 Finally, we have performed all the previous simulations
introducing a small parametric disorder, in all cases the results
are qualitatively similar. These can be summarized in the
following table:

\begin{center}

\begin{tabular}{c c c c c}
\hline \hline & Radial mov. & Twist mov. & Mixed regime. & Mixed
regime.  \\ & regime & regime & Radial pert. & Angular pert. \\
\hline \hline A-T chain & Reflection & Trans. & Reflection &
Trans. \\ \hline C-G chain & Trapping & Trans. & Trapping & Trans.
\\ \hline \hline
\end{tabular}
\label{table}
\end{center}

 The results are clear, the only possible candidate for charge
transport mediated by polarons in an inhomogeneous chain are the
twist modes. Accurate experiments are needed to find out good
values of the parameters and for determining which kind of regime
is possible in DNA and, therefore, if polarons may have a role in
charge transport.

Some preliminary numerical tests with a more realistic model where
the  ionization energy of a C-G base pair is lower than the
A-T base pair by an amount of the order of 0.5 eV, show a
trapping phenomenon, similar as the observed experimentally
\cite{GAKSW01}, where charge moves in an A-T chain by means of
twist modes and reach a C-G base pair. In all other cases we have
always observed a reflection phenomenon. This problem will be the
object of further research.

\section{Conclusions}

 We have considered a fully nonlinear, three--dimensional, semi--classical,
tight--binding model for charge transport in DNA made out of
identical base pairs except for a single one of different type,
which we call a base pair inhomogeneity. There are two types of this
inhomogeneity, soft, composed of G-G base pairs with and A-T inhomogeneity and
hard, with the complementary composition.

In a previous work~\cite{PAHR03}, it has been described that in this system there
exist two types of polarons, twist polarons and radial polarons, for which the
electronic variables are coupled essentially to the radial or angular modes. The
existence of these polarons depends both on the system parameters and on the
form the movement of the polaron is activated.
The properties of the two types of moving polarons are different.
In general, the twist polarons are more robust with
respect to parametric disorder, the polaron has lower
velocity, and the activation energy is higher than in the radial
movability regime.

In an inhomogeneous chain with a single different base pair as a
local inhomogeneity, we have observed that moving polarons
activated by angular perturbations are always transmitted by the
inhomogeneity. If the polarons are activated by radial
perturbations, we have never observed a transmission of the charge
across the inhomogeneity.

Some recent experiments on electron transfer in the DNA molecule
\cite{GAKSW01} show that electrons can migrate over long
distances between a triplet C-G base pairs and a C-G base pair
separated by a number n of A-T base pairs. Moreover, the triplet C-G
base pairs acts as a sink for holes in the chain. In our model,
decreasing slightly the ionization energy of the C-G base pair, as
in real DNA occurs, we are able to reproduce the experimental
results. Some more detailed numerical simulations are currently
underway and will be published in due time. Also, these
experimental techniques could be of application to contrast
experimentally our results.

In our model, we have considered a DNA chain in the vacuum. We
have focused in the polaronic character of the charge carrier and
its interaction with the chain. If the influence of the medium is
not too strong, it is expected that the results would be similar as
the ones obtained in the vacuum. In fact, they  are in agreement with
some experimental observations on DNA in aqueous solutions
\cite{GAKSW01}. The inclusion of thermal effects will be the
subject of further studies, although  in similar systems it has
been shown that the main characteristics of the charge transport do
not change very much when the temperature is considered \cite{Arc03}.
\appendix

\section{Stationary polaron-like states}
 In this section we expose the procedure followed for obtaining
linearly stable, stationary localized states of our model given by
Eqs.~(\ref{electron}-\ref{angular}). These solutions have been
used in the previous section in order to generate mobile polarons.
 Since the adiabaticity parameter $\tau$ is small enough, the fastest
variables are the $\{c_n\}$, with a characteristic frequency (the
linear frequency of the uncoupled system) of order $1/\tau \sim
19$, followed by the $\{r_n\}$ with frequency unity, and the
$\{\phi_n\}$ with $\Omega_{\phi}\sim 0.11$. We can suppose
initially that $r_n$ and $\phi_n$ are constant,i.e., we use the
Born--Oppenheimer approximation. For this purpose, we use a
modification of the numerical method outlined in
Refs.~\cite{KAT98,VT01}. We substitute in Eq.~(\ref{electron})
$c_n=\Phi_n \exp(-\ii\,E\,t/\tau)$, with time--independent
$\Phi_n$'s, and we obtain a nonlinear difference system $E
\Phi=\widehat{A}\Phi$, with $\Phi=(\Phi_1,...,\Phi_N)$, from which
a map $\Phi'=\widehat{A}\Phi/\|\widehat{A}\Phi\|$ is constructed,
$\|.\|$ being the quadratic norm.

 Thus, using Eqs.~(\ref{electron}-\ref{angular}), the stationary
solutions must be  attractors of the  map:
\begin{eqnarray}
 r'_{n}&=&-\frac{k}{b_n}\,|c_n|^2\ \\
& & -\, \frac{\alpha}{b_n}\, \left[\frac{\partial
d_{n,n-1}}{\partial
r_n}(c_{n}^*c_{n-1}+c_{n}c_{n-1}^*)+\frac{\partial
d_{n+1,n}}{\partial r_n}(c_{n+1}^*c_{n}+c_{n+1}c_{n}^*) \right],
\label {radialmap} \\ \nonumber \phi'_n
&=&\frac{1}{2}(\phi_{n+1}+\phi_{n-1})\,
\\  & & -\,
  \frac{\alpha V}{2 \Omega^2}\,\left[\frac{\partial d_{n,n-1}}{\partial \phi_n}
(c_{n}^*c_{n-1}+c_{n}c_{n-1}^*)+ \frac{\partial
d_{n+1,n}}{\partial \phi_n}(c_{n+1}^*c_{n}+c_{n+1}c_{n}^*)\right]
\label {angularmap}, \\
 c'_{n}&=&\frac{[(E_n\,+k\,r'_n)\,c_n
\label{electronmap} -(1-\alpha\,d'_{n+1,n})\,c_{n+1}
-(1-\alpha\,d'_{n\,n-1})\,c_{n-1}]}{\|\{(E_n\,+k\,r'_n)\,c_n
-(1-\alpha\,d'_{n+1,n})\,c_{n+1}
-(1-\alpha\,d'_{n\,n-1})\,c_{n-1}\}\|},
\end{eqnarray}
where $d'=d(r',\phi')$. The starting point is a completely
localized state given by $c_n=\delta_{n,0}$, $r_n=0$ and
$\phi_n=0$, $\forall n$. Then the map is applied until convergence
is achieved. In this way both stationary solutions and their
energy $E$ can be obtained.

\begin{figure}
\begin{center}
\includegraphics[width=\singlefig]{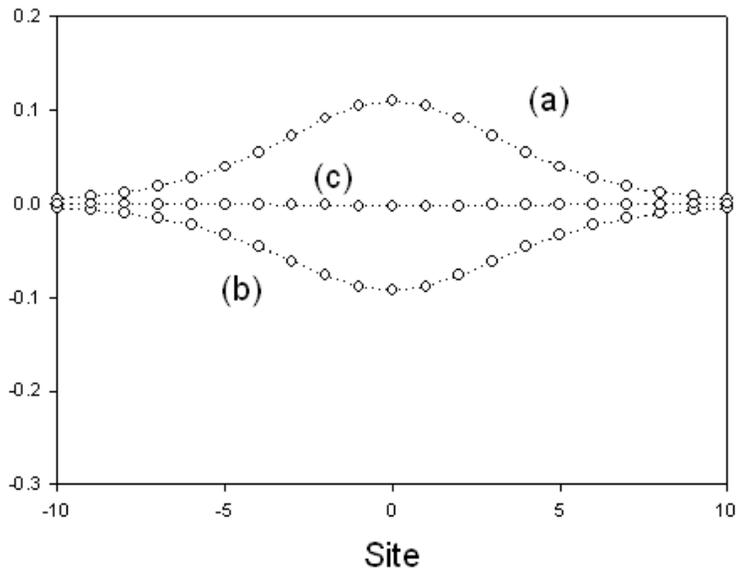} 
\end{center}
\caption{Profiles of the ground state in the homogeneous and
ordered chain. (a) Wave function amplitudes $|c_n|^2$. (b) Static
radial displacements $r_n$. (c) Static twists elongations
$\theta_{nn-1}$}
\label{gsho}
\end{figure}

 Firstly, we have analyzed the homogeneous and ordered case , i.e.,
$E_n^0=E_0$, that arises in synthetic DNA (the constant $E_0$ can
be set to zero by means of a gauge transformation). As shown in
Fig.~\ref{gsho}, in a typical ground state the charge is fairly
localized at few sites, and the amplitudes decay monotonically and
exponentially with growing distance from the central site. The
associated patterns of the static radial and relative angular
displacements are similar

\begin{figure}
\begin{center}
\includegraphics[width=\singlefig]{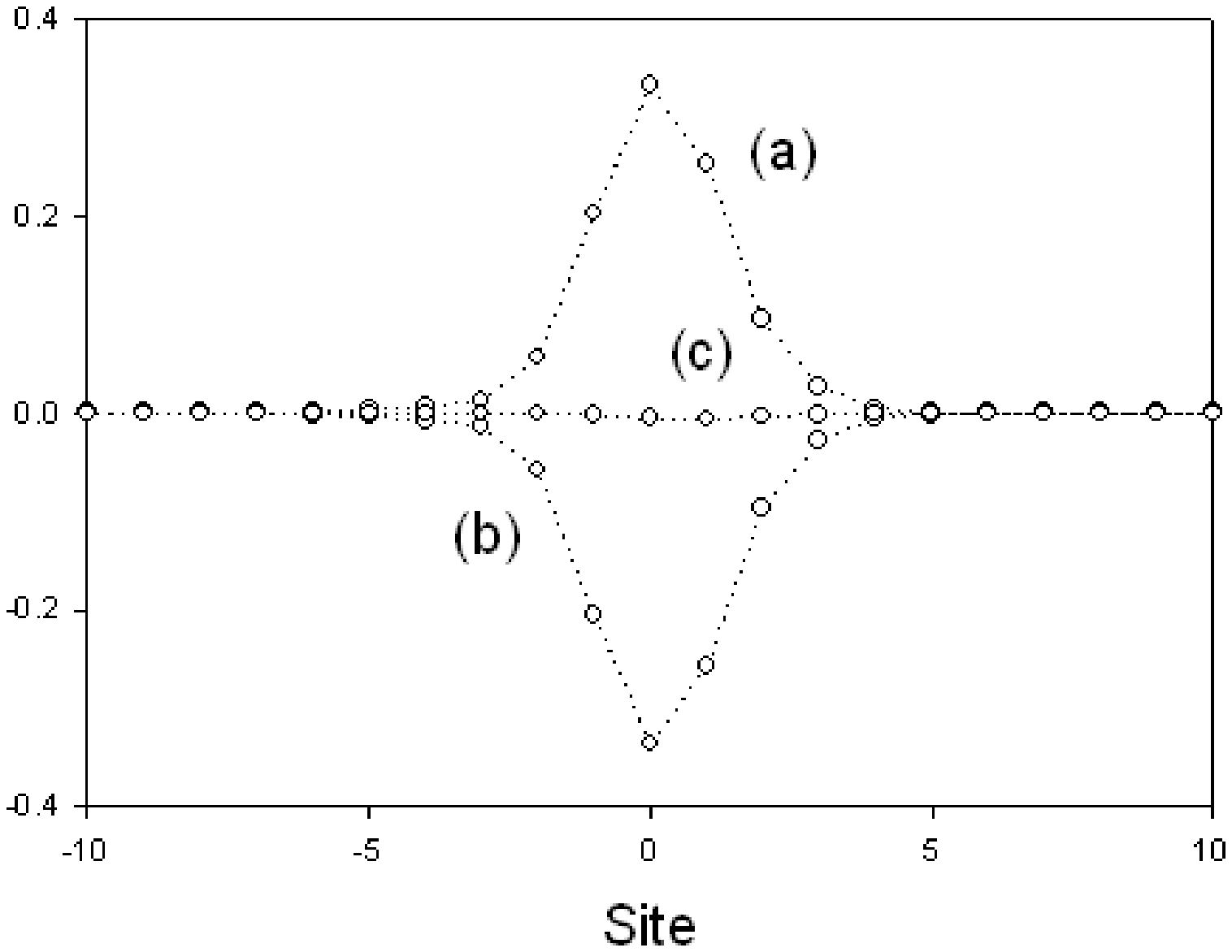} 
\end{center}
\caption{Profiles of the ground state in the homogeneous and
disordered chain. (a) Wave function amplitudes $|C_n|^2$. (b)
Static radial displacements $r_n$. (c) Static twists elongations
$\theta_{nn-1}$}
\label{gshd}
\end{figure}

 We can introduce a certain degree of parametric disorder in the
on--site electronic energy $E_n^0$ by means of a random potential
$E_n^0\in [-\Delta E,\Delta E]$, with mean value zero and
different interval sizes $\Delta E$. In this case, the localized
excitation patterns do not change qualitatively, as  shown in
Fig.~\ref{gshd}. However, as the translational invariance is broken by
the disorder, the localized excitation pattern is not symmetric
with respect to a lattice site, which is different to the ordered
case. Also, the localization is enhanced with the disorder, due to
Anderson localization~\cite{Anderson58}.

\begin{figure}
\begin{center}
\includegraphics[width=\singlefig]{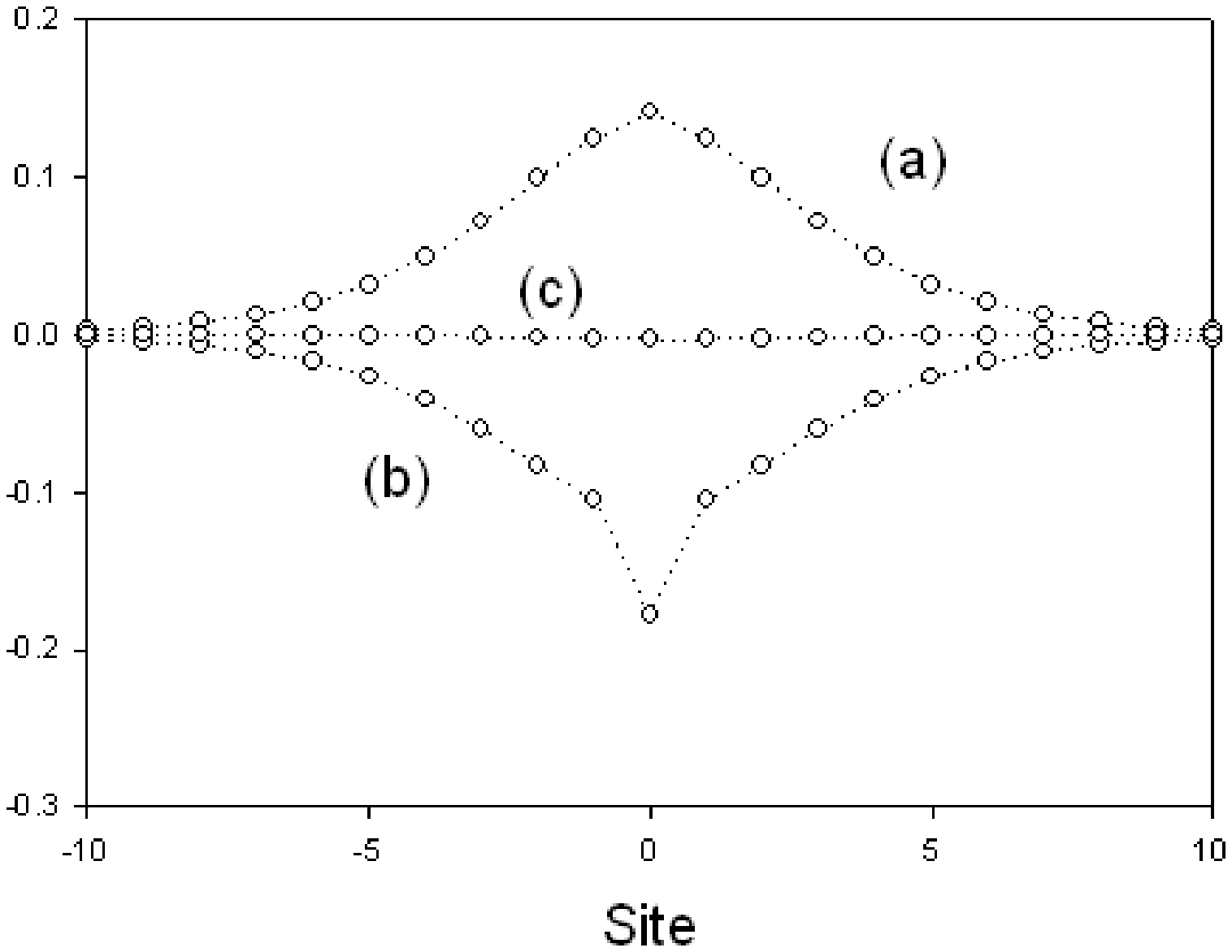}
\end{center}
\caption{Profiles of the ground state in a C-G chain centered in
the inhomogeneity, an A-T base pair. (a) Wave function amplitudes
$|C_n|^2$. (b) Static radial displacements $r_n$. (c) Static
twists elongations $\theta_{n\,n-1}$}
 \label{gsio1}
\end{figure}

\begin{figure}
\begin{center}
\includegraphics[width=\singlefig]{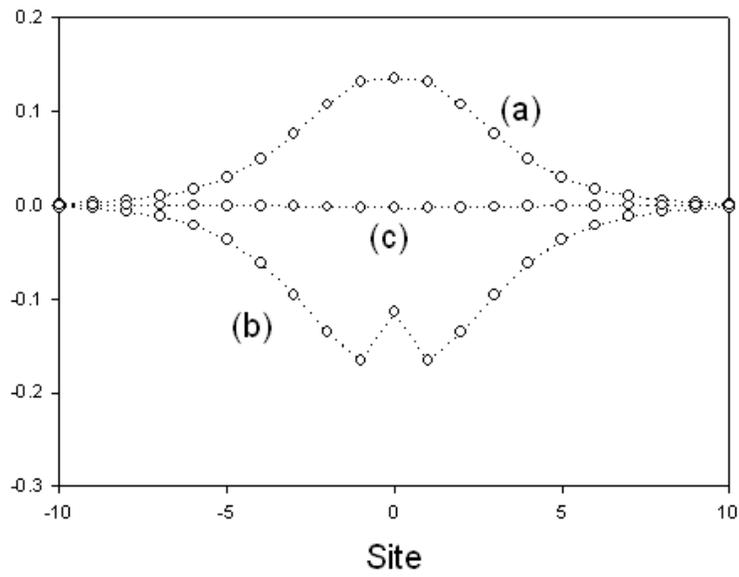} 
\end{center}
\caption{Profiles of the ground state in an A-T chain centered in
the inhomogeneity, a C-G base pair.(a) Wave function amplitudes
$|C_n|^2$. (b) Static radial displacements $r_n$. (c) Static
twists elongations $\theta_{n\,n-1}$}
\label{gsio2}
\end{figure}

 We consider now a chain with a local inhomogeneity due to the
existence of a single base pair which is different to the other
ones (without disorder). Our results show that the ground state
centered in a site far from this local inhomogeneity is
qualitatively similar to the obtained in the homogeneous chain
case. Nevertheless, differences appear if we consider the
stationary state centered in the inhomogeneity. In a C-G chain,
with an A-T base pair as the local inhomogeneity, the shape of the
ground state is qualitatively similar to the obtained in the
homogeneous case, as shown in Fig.~\ref{gsio1}. In an A-T chain,
with a C-G base pair as the local inhomogeneity, the static radial
displacements are different, as shown in Fig.~\ref{gsio2}. As we
have seen in the previous section, this fact is determinant in
relation with the transmission, reflection or trapping of a moving
polaron by the inhomogeneity.


\section*{Acknowledgments}
 The authors acknowledge Dr. Jose Mar\'{\i}a Romero, from the GFNL of the University of Sevilla,
 for valuable suggestions. They are grateful to partial support under the LOCNET EU network
HPRN-CT-1999-00163. JFRA  acknowledges DH and the
 Institut f\"{u}r Theoretische Physik for their warm hospitality.
D.H. acknowledges support by the
Deutsche Forschungsgemeinschaft via a Heisenberg fellowship
(He 3049/1-1).

\newcommand{\noopsort}[1]{} \newcommand{\printfirst}[2]{#1}
  \newcommand{\singleletter}[1]{#1} \newcommand{\switchargs}[2]{#2#1}

\end{document}